\documentclass[a4paper]{amsart}
\usepackage{amsmath,amsthm,amssymb}
\usepackage{epic,eepic}
\usepackage{epsfig,cite,indentfirst,oldgerm}

\begin{document}

\title[XXZ and KP]
{XXZ scalar products and KP}

\author{O Foda, M Wheeler and M Zuparic}

\address{Department of Mathematics and Statistics,
         University of Melbourne,
         Parkville, Victoria 3010, Australia.}
\email{foda, mwheeler, mzup@ms.unimelb.edu.au}

\keywords{XXZ spin chain. KP. Free fermions}
\subjclass[2000]{Primary 82B20, 82B23}

\begin{abstract}
Using a Jacobi-Trudi-type identity, we show that the scalar product
of a general state and a Bethe eigenstate in a finite-length XXZ
spin-$\frac{1}{2}$ chain is (a restriction of) a KP $\tau$ function.
This leads to a correspondence between the eigenstates and points
on Sato's Grassmannian. Each of these points is a function of
the rapidities of the corresponding eigenstate, the
inhomogeneity variables of the spin chain and the crossing
parameter.

\end{abstract}

\maketitle

\def\ll{\left\lgroup}
\def\rr{\right\rgroup}

\def\proofend{\ensuremath{\square}}
\def\det{\operatorname{det}}
\def\psis{\psi^{*}}
\def\Psis{\Psi^{*}}
\def\lprod{\mathop{\prod{\mkern-29.5mu}{\mathbf\longleftarrow}}}
\def\rprod{\mathop{\prod{\mkern-28.0mu}{\mathbf\longrightarrow}}}
\def\complex{\mathbb{C}}
\def\integer{\mathbb{Z}}
\def\no{\nonumber}

\newcommand{\field}[1]{\mathbb{#1}}
\newcommand{\C}{\field{C}}
\newcommand{\N}{\field{N}}
\newcommand{\Z}{\field{Z}}
\newcommand{\R}{\field{R}}
\newtheorem{Theorem}{Theorem}[section]
\newtheorem{Corollary}[Theorem]{Corollary}
\newtheorem{Proposition}[Theorem]{Proposition}
\newtheorem{Conjecture}[Theorem]{Conjecture}
\newtheorem{Lemma}[Theorem]{Lemma}
\newtheorem{Example}[Theorem]{Example}
\newtheorem{Note}[Theorem]{Note}
\newtheorem{Definition}[Theorem]{Definition}
\newtheorem{ca}{Figure}

\setcounter{section}{-1}

\section{Introduction}
\label{introduction}

In \cite{fwz} we observed that the partition function of the six
vertex model on a finite-size square lattice with domain wall
boundary conditions, $Z_N$, is (a restriction of) a KP $\tau$
function, for all values of the crossing parameter\footnote{We
refer to \cite{blue-book} for an introduction to classical
integrable models and KP theory, and to \cite{korepin-book}
for an introduction to quantum integrable models and the
algebraic Bethe Ansatz.}.

In this work, we extend the above result as follows.
We prove a Jacobi-Trudi-type identity that implies that
certain determinants are restrictions of KP $\tau$ functions.
Next, for a length-$M$ XXZ spin-$\frac{1}{2}$ chain,
we consider the scalar product
$\langle \{ \lambda \}|\{ \mu\}_\beta \rangle$,
where $\langle \{ \lambda \}|$ is a general state,
$\{ \lambda\}$ are $N$ free variables,
$|\{ \mu\}_\beta \rangle$ is a Bethe eigenstate,
$\{ \mu \}_{\beta}$ are $N$ variables that
satisfy the Bethe equations, and $N\leq M$.
We use the Jacobi-Trudi-type identity to show that
$\langle \{ \lambda \}|\{ \mu\}_\beta \rangle$
is a restricted KP $\tau$ function, where the 
(infinitely many) KP time variables are functions 
of the $N$ free variables $\{ \lambda\}$.
We obtain an expression for
$\langle \{\lambda \}|\{ \mu\}_\beta \rangle$
as an expectation value of charged free fermions,
then {\it \lq peel off\rq\,} the time dependencies 
to obtain a correspondence between each eigenstate
$|\{\mu\}_\beta \rangle$
and a point on Sato's Grassmannian represented by
the action of exponentials of fermion bilinears.
The coefficients of the bilinears are functions
of $\{\mu\}_{\beta}$, the inhomogeneities of the 
spin chain and the crossing parameter.

In section {\bf 1}, we recall basic definitions related to
symmetric functions and to KP $\tau$ functions. From a KP
$\tau$ function with no constraints on the time variables
we obtain a {\it restricted} KP $\tau$ function by setting
the (infinitely many) KP time variables to be power sums in
$N$ independent variables.
In {\bf 2}, we prove a Jacobi-Trudi-type identity which
implies that certain determinants are restricted KP $\tau$
functions.  In {\bf 3}, we recall basic facts related to
the algebraic Bethe Ansatz approach to the XXZ
spin-$\frac{1}{2}$ chain, and particularly Slavnov's
determinant expression for the scalar product of a general
state and a Bethe eigenstate \cite{slavnov}.

In section {\bf 4}, we show that the Jacobi-Trudi-type identity
introduced in section {\bf 2} implies that Slavnov's determinant
is a restricted KP $\tau$ function. The result of \cite{fwz}, 
that the domain wall partition function $Z_N$ is a restricted 
KP $\tau$ function follows as a special case.
In {\bf 5}, we write Slavnov's determinant as an expectation
value of charged free fermions. In {\bf 6}, we propose a correspondence
between the Bethe eigenstates and points on Sato's Grassmannian,
and in {\bf 7}, we include a number of remarks.

\section{Symmetric functions and KP $\tau$ functions}
\label{functions}

\subsection{Elementary symmetric functions} Following \cite{macdonald}, 
the {\it elementary symmetric function} $e_i\{x\}$ in $N$ variables 
$\{x\}$ is the $i$-th coefficient in the generating series

\begin{equation}
\sum_{i=0}^{\infty}
e_i\{x\}
\, 
k^i
=
\prod_{i=1}^{N}
\left( 1 + x_i \, k \right)
\label{e-s}
\end{equation}

For example, $e_0\{x\} = 1$, $e_1(x_1,x_2,x_3) = x_1+x_2+x_3$,
$e_2(x_1,x_2) = x_1x_2$. $e_i\{x\} = 0$, for $i<0$ and for 
$i>N$.

\subsection{Complete symmetric functions} The {\it complete
symmetric function} $h_i\{x\}$ in $N$ variables $\{x\}$ is
the $i$-th coefficient in the generating series

\begin{equation}
\sum_{i=0}^{\infty}
h_i\{x\}
\, 
k^i
=
\prod_{i=1}^{N}
\frac{1}{1-x_i \, k}
\label{c-s}
\end{equation}

For example, $h_0\{x\} = 1$, $h_1(x_1,x_2,x_3) = x_1+x_2+x_3$,
$h_2(x_1,x_2) = x_1^2+x_1x_2+x_2^2$, and 
$h_{i}\{x\} = 0$ for $i<0$. In the sequel, we will use
the identities

\begin{eqnarray}
h_i
\{\widehat{x}_m \} + x_m h_{i-1}\{x\} &=& h_i\{x\}
\label{i1}
\\
h_i\{\widehat{x}_l\}-h_i\{\widehat{x}_m\}&=& (x_m-x_l) \, h_{i-1}\{x\}
\label{i2}
\end{eqnarray}

\noindent where $\{\widehat{x}_m\}$ is the set of variables $\{x\}$,
with the omission of $x_m$. Equation (\ref{i1}) follows from 
(\ref{c-s}), and (\ref{i2}) is a re-arrangement of (\ref{i1}). 

\subsection{Schur functions}

\noindent The {\it Schur function} $s_{\lambda}\{x\}$ indexed by
a Young diagram $\lambda=[\lambda_1,\ldots,\lambda_r]$ with $r$ 
non-zero-length rows, $r \leq N$, is

\begin{equation}
s_{\lambda}\{x\}
=
\frac{
\det
\ll
x_i^{\lambda_j-j+N}
\rr_{1 \leq i,j \leq N}
}
{\prod_{1\leq i<j \leq N} \left( x_i-x_j \right)}
=
\det
\ll
h_{\lambda_i-i+j}\{x\}
\rr_{1\leq i,j \leq N}
\label{schur}
\end{equation}

\noindent where $\lambda_{i} =  0$, for 
$r+1 \leq i \leq N$. For example, $s_{\phi} \{x\} =1$,
$s_{[1  ]}(x_1,x_2,x_3) = x_1+x_2+x_3$,
$s_{[1,1]}(x_1,x_2    ) = x_1 x_2$.
The first and second equalities in Equation (\ref{schur}) 
are the definition and the {\it Jacobi-Trudi identity} for 
the Schur functions, $s_{\lambda}\{x\}$, which form a basis 
for the ring of symmetric functions in $\{x\}$.

\subsection{Character polynomials}

The {\it one-row character polynomial} $\chi_{i}\{t\}$
is the $i$-th coefficient in the generating series

\begin{equation}
\sum_{i=0}^{\infty}
\chi_{i}\{t\}
\, 
k^i
=
\exp
\ll
\sum_{i=1}^{\infty}
t_i \, k^i
\rr
\label{one-row-character}
\end{equation}

For example,
$\chi_0\{t\} = 1$,
$\chi_1\{t\} = t_1$,
$\chi_2\{t\} = \frac{t_1^2}{2}+t_2$,
$\chi_3\{t\} = \frac{t_1^3}{6}+ t_1 t_2 + t_3$,
and $\chi_{i}\{t\} = 0$ for $i<0$.
The {\it character polynomial} $\chi_{\lambda}\{t\}$ 
indexed by a Young diagram 
$\lambda = [\lambda_1, \lambda_2, \ldots, \lambda_r]$, 
with $r$ non-zero-length rows, $r \leq N$, is

\begin{equation}
\chi_{\lambda}\{t\}
=
\det
\ll
\chi_{\lambda_i-i+j}\{t\}
\rr_{1\leq i,j \leq n}
\label{r-row-character}
\end{equation}

For example
$\chi_{[1,1]}\{t\} = \frac{t_1^2}{2}-t_2$,
$\chi_{[2,1]}\{t\} = \frac{t_1^3}{3}-t_3$.
Notice that $\chi_{\lambda}\{t\}$ is a polynomial in infinitely many
variables $\{t\}$, in the sense that it can depend on all $t_i$, for
$i \leq |\lambda|$, where $|\lambda|$ is the sum of the lengths
of all rows in $\lambda$. As $|\lambda| \rightarrow \infty$, more time
variables $t_i$ contribute to $\chi_{\lambda}\{t\}$, which 
form a basis for the ring of symmetric
functions in the infinitely-many variables $\{t\}$.

\subsection{From character polynomials to Schur functions}
The restriction 
$ t_m        \rightarrow \frac{1}{m} \sum_{i=1}^{N} x_i^m$
sends
$\chi_i\{t\} \rightarrow                h_i \{x\}$.
Consequently, the character polynomials 
$\chi_{\lambda} \{t\}$ in infinitely many time variables $\{t\}$, 
where $\lambda$ has $r$ non-zero-length rows, become equal to 
the Schur functions $s_{\lambda} \{x\}$ in $N$ variables $\{x\}$, 
where $N \geq r$, by setting
$t_m \rightarrow \frac{1}{m}\sum_{i=1}^{N} x_i^m$, for $m \geq 1$.

\subsection{KP $\tau$ functions}

A function $\tau\{t\}$ in the infinitely-many time variables 
$\{ t\}$ is a KP $\tau$ function if and only
if it can be expanded in the basis of character polynomials
$\chi_{\lambda}\{t\}$ as

\begin{equation}
\tau\{t\} = \sum_{\lambda} c_{\lambda} \, \chi_{\lambda}\{t\}
\label{tau}
\end{equation}

\noindent where the coefficients $c_{\lambda}$ satisfy 
Pl\"ucker relations \cite{blue-book, hirota}.

\subsection{Restricted KP $\tau$ functions}

We define a {\it restricted} $\tau$ function $\tau\{x\}$ to be
a $\tau$ function $\tau\{t\}$ whose infinitely many time
variables $\{t\}$ have been set to
$t_m \rightarrow \frac{1}{m}\sum_{i=1}^{N} x_i^m$ for $m\geq 1$,
that is, power sums in the $N$ variables $\{x\}$. A restricted
$\tau$ function is symmetric in the $N$ variables $\{x\}$,
since from (\ref{tau}) it has the form

\begin{equation}
\tau\{x\} = \sum_{\lambda \subseteq [(M-1)^{N}]} 
c_{\lambda} 
\, 
s_{\lambda} \{ x \}
\label{tau-res}
\end{equation}

\noindent where the coefficients $c_{\lambda}$ satisfy Pl\"ucker
relations, and $M\geq 1$. On the right hand side of Equation 
(\ref{tau-res}), we indicated that the sum is over all Young 
diagrams $\lambda \subseteq [(M-1)^{N}]$. That the maximal number 
of allowed rows in $\lambda$ is $N$ is due to the fact that 
there are only $N$ independent variables $\{x\}$. That the 
number of columns is $M-1$, is due to the fact that in this
work, we consider only polynomials in $\{x\}$. The precise 
value of $M$ is at this stage unspecified. In the sequel, 
$\tau\{t\}$ functions of the infinitely many variables $\{t\}$ 
are unrestricted, while the $\tau\{x\}$ functions of the $N$ 
variables $\{x\}$ are restricted.

\subsection{Determinants that are restricted KP $\tau$ functions}

There are determinants that can be put in the form on the right 
hand side of Equation (\ref{tau-res}), hence they are restricted 
KP $\tau$ function, as shown in the following lemma.

\bigskip
\noindent
{\bf Lemma 1.}
\ \ \
Let $\mathcal{H}$ be a matrix with entries
$\mathcal{H}_{ij}=h_{j-i}\{x\}$, 
$1\leq i \leq N$, 
$1\leq j \leq N+M-1$,
where $\{x\}$ is a set of $N$ variables.
Let $\mathcal{C}$ be a constant matrix with entries
$\mathcal{C}_{ij}=c_{ij}$, 
$1 \leq i \leq N+M-1$ and
$1 \leq i \leq N$.
We choose $M \geq 1$.
The product $\mathcal{H}\mathcal{C}$ is an $N\times N$ 
matrix, 
and $\det \ll \mathcal{H}\, \mathcal{C} \rr$ is a restricted
$\tau$ function of the KP hierarchy.

\medskip
\noindent
{\bf Proof.} In the following calculation, all determinants
are over $N\times N$ matrices, so we do not need to write 
the ranges of the row and column indices explicitly.

\begin{eqnarray}
\label{product-form}
\\
\det
\ll
\mathcal{H}\, \mathcal{C}
\rr
&=&
\det
\ll
\sum_{k=1}^{N+M-1} h_{k-i}\{x\} \, c_{kj}
\rr
\nonumber
\\
&=&
\sum_{1 \leq k_1 < \cdots < k_N \leq N+M-1}
\det
\ll
h_{k_j-i}\{x\}
\rr
\det \ll c_{k_i,j} \rr
\nonumber
\\
&=&
\sum_{0 \leq \lambda_N \leq \cdots \leq \lambda_1 \leq M-1}
\det
\ll
h_{\lambda_{(N-j+1)}+j-i} \{x\}
\rr
\det
\ll
c_{\lambda_{(N-i+1)}+i, j}
\rr
\nonumber
\\
&=&
\sum_{0 \leq \lambda_N \leq \cdots \leq \lambda_1 \leq M-1}
\det
\ll
h_{\lambda_{i}-i+j}\{x\}
\rr
\det
\ll
c_{\lambda_{(N-i+1)}+i,j}
\rr
\nonumber
\\
&=&
\sum_{\lambda \subseteq [(M-1)^{N}]}
c_{\lambda} \, s_{\lambda} \{x\}
\nonumber
\end{eqnarray}

\noindent where
$c_{\lambda} =
\det
\ll
c_{(\lambda_{(N-i+1)}+i),j}
\rr
$.
By construction, the coefficients $c_{\lambda}$ obey Pl\"ucker 
relations \cite{blue-book, hirota}, and the required form 
(\ref{tau-res}) is obtained.

\section{A Jacobi-Trudi-type identity}
\label{jacobi-trudi}

In the following, we show that an expression that shows up frequently 
in the theory of quantum integrable models can be put in the form of 
the left hand side of Equation (\ref{product-form}), hence it is also
a restricted KP $\tau$ function. 

\bigskip
\noindent
{\bf Lemma 2.} Let $\kappa$ be an $(N+M-1)\times N$ matrix with 
coefficients $\kappa_{ij}$ that do not depend on $\{x\}$, and 
$M \geq 1$. Then

\begin{equation}
\frac{
\det
\ll
\sum_{k=1}^{N+M-1} x_i^{k-1} \kappa_{kj}
\rr_{1 \leq i, j \leq N}}
{\prod_{1 \leq i< j \leq N} \left(x_i -x_j \right)}
=
\det
\ll
\sum_{k=1}^{N+M-1}
h_{k-i} \{x\}
\,
\kappa_{k, N-j+1}
\rr_{1 \leq i, j \leq N}
\label{lemma-matrix}
\end{equation}

\noindent {\bf Proof.} The first step in the proof is to cancel 
the factor in the denominator of the left hand side of Equation 
(\ref{lemma-matrix}). We define the operator $\mathcal{R}_{i}$ 
which acts on the $i$-th row of an arbitrary $N\times N$ matrix 
$\mathcal{A}$. In component form, we write

\begin{equation}
\ll
\mathcal{R}_{i'} \mathcal{A}
\rr_{ij}
=
\mathcal{A}_{ij} - \delta_{ii'} \mathcal{A}_{i+1, j}
\label{row-op1}
\end{equation}

Let $\mathcal{M}^{(1)}$ be an $N\times N$ matrix with components
$\mathcal{M}^{(1)}_{ij} = \sum_{k=1}^{N+M-1} \, x_i^{k-1} \, \kappa_{kj}$.
Observing that 
$\mathcal{M}^{(1)}_{ij} =
 \sum_{k=1}^{N+M-1} \, h_{k-1}(x_i) \, \kappa_{kj}$, and making repeated 
use of Equation (\ref{i2}), we find the string of relations

\begin{eqnarray}
\det
\ll
\prod_{i=1}^{N-1}
\mathcal{R}_{i}
\, 
\mathcal{M}^{(1)}
\rr
&=&
\det \mathcal{M}^{(2)}
\prod_{i=1}^{N-1} \ll x_i-x_{i+1} \rr
\label{string}
\\
\det
\ll
\prod_{i=1}^{N-2}
\mathcal{R}_{i}
\, 
\mathcal{M}^{(2)}
\rr
&=&
\det \mathcal{M}^{(3)}
\prod_{i=1}^{N-2} \ll x_i-x_{i+2} \rr
\nonumber
\\
&\vdots&
\nonumber
\\
\det
\ll
\mathcal{R}_{1}
\, 
\mathcal{M}^{(N-1)}
\rr
&=&
\det \mathcal{M}^{(N)}
\ll x_1-x_{N} \rr
\nonumber
\end{eqnarray}

\noindent where
$\prod_{i=1}^{n}\mathcal{R}_i = \mathcal{R}_n \ldots \mathcal{R}_1$.
For simplicity, we show only the components of the last determinant

\begin{equation}
\mathcal{M}^{(N)}_{ij}
=
\sum_{k=1}^{N+M-1} h_{k-N+i-1} \{x\}_{i} \, \kappa_{kj}
\label{lemma-matrix2}
\end{equation}

\noindent where $\{x\}_i$ is the set of $N-i+1$ variables
$x_i, \ldots, x_N$. Due to invariance of the determinant 
under such row operations, we can combine the string of 
relations in Equation (\ref{string}) into a single equation

\begin{equation}
\frac{
\det
\mathcal{M}^{(1)}
}{
\prod_{1\leq i< j \leq N} \left( x_i - x_j \right)
}
=
\det \mathcal{M}^{(N)}
\label{lemma-matrix3}
\end{equation}

We have almost succeeded in obtaining the right-hand-side of Equation
(\ref{lemma-matrix}), except that the complete symmetric functions in
$\mathcal{M}_{ij}^{(N)}$, as shown in Equation 
(\ref{lemma-matrix2}) depend on sets of variables of different
cardinalities.

The second step in the proof is to use row operations to introduce
dependence on the full set of $N$ variables $\{x\}$ in each of the 
complete symmetric
functions. Introduce the operator $\mathcal{R}_{i}(x)$, which differs
from that used in the first step in that it depends on a (single) 
variable $x$ and acts on the $i$-th row of an arbitrary matrix 
$\mathcal{A}$ to give

\begin{equation}
\ll \mathcal{R}_{i'}(x) \, \mathcal{A} \rr_{ij}
=
\mathcal{A}_{ij} + \delta_{ii'} \, x \, \mathcal{A}_{i-1,j}
\label{row-op2}
\end{equation}

Using Equation (\ref{lemma-matrix2}) for $\mathcal{M}_{ij}^{(N)}$ and
Equation (\ref{i1}) repeatedly, we obtain another string of relations

\begin{eqnarray}
\det
\ll
\prod_{i=N}^{2}
\mathcal{R}_{i}(x_{i-1})
\, 
\mathcal{M}^{(N)}
\rr
&=&
\det
\mathcal{M}^{(N+1)}
\label{string2}
\\
\det
\ll
\prod_{i=N}^{3}
\mathcal{R}_{i}(x_{i-2})
\, 
\mathcal{M}^{(N+1)}
\rr
&=&
\det
\mathcal{M}^{(N+2)}
\nonumber
\\
&\vdots&
\nonumber
\\
\det
\ll
\mathcal{R}_{N}(x_1)
\, 
\mathcal{M}^{(2N-2)}
\rr
&=&
\det
\mathcal{M}^{(2N-1)}
\nonumber
\end{eqnarray}

\noindent where again, for simplicity, we show only the components
of the final determinant

\begin{equation}
\mathcal{M}^{(2N-1)}_{ij}
=
\sum_{k=1}^{N+M-1} h_{k-N+i-1}\{x\} \, \kappa_{kj}
\end{equation}

\noindent and the complete symmetric functions now depend on the
full set of $N$ variables $\{x\}$. Again, since determinants are 
invariant under these row operations, the string of relations 
in Equation (\ref{string2}) can be combined into the single equation

\begin{equation}
\det{\mathcal{M}^{(N)}}
=
\det\mathcal{M}^{(2N-1)}
=
\det
\ll
\sum_{k=1}^{N+M-1}
h_{k-i} \{x\} 
\,
\kappa_{k,N-j+1}
\rr
\label{anotherequation}
\end{equation}

\noindent where we have performed trivial index shifts in
$\det \mathcal{M}^{(2N-1)}$ to obtain the right hand side
of Equation (\ref{anotherequation}). Recalling Equation 
(\ref{lemma-matrix3}), the proof is complete. The classical 
Jacobi-Trudi identity is recovered from Equation 
(\ref{lemma-matrix}) by setting $M-1 = \lambda_1$, 
and $\kappa_{kj} = \delta_{k-1, \lambda_j - j +N}$.

\subsection{Corollary} Writing $c_{kj} = \kappa_{k,N-j+1}$, 
it follows that the left hand side of Equation (10) is identical
to the right hand side of Equation (11), and that any expression 
in the form of the left hand side of Equation (11) is a restricted 
KP $\tau$ function.

\section{XXZ and the algebraic Bethe Ansatz}
\label{xxz-alg-bethe}

\subsection{The XXZ Hamiltonian}

Let $\sigma_i^{x,y,z}$ be Pauli matrices acting in space $V_i$
isomorphic to $\mathbb{C}^2$. The Hamiltonian of a length-$M$
XXZ spin-$\frac{1}{2}$ chain is

\begin{equation}
H
=
\sum_{i=1}^{M}
\ll
\sigma_i^x \, \sigma_{i+1}^x
+
\sigma_i^y \, \sigma_{i+1}^y
+
\Delta
\,
\sigma_i^z \, \sigma_{i+1}^z
\rr
\label{hamiltonian}
\end{equation}

\noindent where periodicity $\sigma_{M+1}^{x,y,z}=\sigma_1^{x,y,z}$ 
is imposed. The eigenstates
and eigenvalues of $H$ are obtained using the algebraic Bethe Ansatz,
as we outline below.

\subsection{The $R$-matrix} Consider the $R$-matrix

\begin{equation}
R_{ab}(\lambda, \mu)
=
\ll
\begin{array}{cccc}
[\lambda - \mu + \gamma] & 0 & 0 & 0 \\
0 & [\lambda - \mu] & [\gamma]   & 0 \\
0 & [\gamma] & [\lambda - \mu]   & 0 \\
0 & 0 & 0 & [\lambda - \mu + \gamma]
\end{array}
\rr_{ab}
\label{Rmat}
\end{equation}

\noindent where $[x] = e^x-e^{-x}$. The subscripts of $R_{ab}$
indicate that it acts in the tensor product $V_a \otimes V_b$, 
where each $V_i$ is isomorphic to $\mathbb{C}^2$.
$R_{ab}$ satisfies the Yang-Baxter equation in
$V_a \otimes V_b \otimes V_c$

\begin{equation}
R_{ab}(\lambda,\mu)\, R_{ac}(\lambda,\nu)\, R_{bc}(\mu,\nu)
=
R_{bc}(\mu,    \nu)\, R_{ac}(\lambda,\nu)\, R_{ab}(\lambda,\mu)
\label{YB}
\end{equation}

The $R$-matrix in Equation (\ref{Rmat}) plays an essential role 
in the algebraic Bethe Ansatz approach for a variety of quantum 
integrable models \cite{korepin-book}, one of which is the XXZ 
spin-$\frac{1}{2}$ chain.

\subsection{The $L$-operator}

To solve a quantum integrable model using the algebraic Bethe
Ansatz, it is necessary to specify the $L$-operator, which is 
specific to the model under consideration. For the XXZ 
spin-$\frac{1}{2}$ chain, the $L$-operator is 
$L_{ab}(\lambda,\nu) = R_{ab}(\lambda,\nu)$. Using the 
Yang-Baxter equation (\ref{YB}), the $L$-operator satisfies 
the local intertwining relation

\begin{equation}
R_{ab}(\lambda,\mu) \, L_{ac}(\lambda,\nu) \, L_{bc}(\mu,\nu)
=
L_{bc}(\mu,    \nu) \, L_{ac}(\lambda,\nu) \, R_{ab}(\lambda,\mu)
\label{intertwining-L}
\end{equation}

\subsection{The monodromy $T$-matrix}
The monodromy matrix $T_{a}$ is defined by

\begin{equation}
T_a(\lambda)
=
\ll
\begin{array}{cc}
A(\lambda) & B(\lambda)
\\
C(\lambda) & D(\lambda)
\end{array}
\rr_{a}
=
L_{a1}(\lambda, \nu_1) \ldots L_{aM}(\lambda,\nu_M)
\label{monodromy}
\end{equation}

\noindent where it is conventional to suppress dependence on the
inhomogeneities $\nu_i$ in $T_a$, and it is implicit
that each of the operators $A$, $B$, $C$, and $D$ acts in the
tensor product $V_1 \otimes \cdots \otimes V_M$. Using Equation
(\ref{intertwining-L}) inductively, one derives the intertwining
relation

\begin{equation}
R_{ab}(\lambda,\mu) \, T_{a}(\lambda) \, T_{b}(\mu)
=
T_{b}(\mu) \, T_{a}(\lambda) \, R_{ab}(\lambda,\mu)
\label{intertwining}
\end{equation}

\noindent which contains all the algebraic relations between
the operators $A$, $B$, $C$, and $D$.

\subsection{The transfer $\mathcal{T}$-matrix} 
The trace of the monodromy matrix 
${\rm Tr}_a T_a(\lambda)$ $=$ 
$\mathcal{T}$              $=$ 
$A(\lambda)+D(\lambda)$, 
is the {\it transfer matrix} of the spin chain.

\subsection{The Bethe equations}

%Using the commutation relations (\ref{intertwining}), it is possible 
%to show that $|\{\mu\}\rangle$ satisfies the eigenvalue equation
%(\ref{eigenstate}) if and only if the parameters $\{\mu\}$ obey
%the Bethe equations

Given a set of $N$ rapidities $\{\mu\}$, the Bethe equations are 
the set of equations
\begin{equation}
(-)^{N-1}
\frac{a(\mu_i)}{d(\mu_i)}
\prod_{j \not= i}^{N}
\frac{[\mu_j - \mu_i + \gamma]}{[\mu_i - \mu_j + \gamma]}
=
1
\label{bethe2}
\end{equation}
\noindent for $1\leq i \leq N$, where the functions $a(\mu)$
and $d(\mu)$ are the respective eigenvalues of the $A(\mu)$ 
and $D(\mu)$ operators
\begin{eqnarray}
A(\mu) |0\rangle = a(\mu) |0\rangle, \ \
D(\mu) |0\rangle = d(\mu) |0\rangle
\end{eqnarray}
The explicit form of $a(\mu)$ and $d(\mu)$ is specific to the
model under consideration. In the XXZ spin chain,
$a(\mu) = \prod_{i=1}^{M} [\mu-\nu_i+\gamma] $,
$d(\mu) = \prod_{i=1}^{M} [\mu-\nu_i] $. 

The Bethe equations are necessary to show that certain states 
are eigenvectors of the transfer matrix $\mathcal{T}$, and the 
Hamiltonian $H$, of the spin chain.

\subsection{The Bethe eigenstates}

The eigenstates of the transfer matrix 
${\rm Tr}_a T_a(\lambda) = A(\lambda)+D(\lambda)$ are also
eigenstates of the Hamiltonian \cite{korepin-book}. The problem
of finding eigenstates of the XXZ Hamiltonian is that of finding
states $|\Psi\rangle$ that satisfy

\begin{equation}
\ll
A(\lambda)+D(\lambda)
\rr
|\Psi\rangle =
\kappa_{\Psi}(\lambda) |\Psi\rangle
\label{eigenstate}
\end{equation}

\noindent for some eigenvalue $\kappa_{\Psi}(\lambda)$. The algebraic
Bethe Ansatz for the eigenstates is

\begin{equation}
|\Psi\rangle = |\{\mu\}_{\beta} \rangle = 
B(\mu_1) \ldots B(\mu_N) |0\rangle
\label{bethe1}
\end{equation}

\noindent where $N \leq M$, the reference state
$|0\rangle = \otimes^{M} \left({1\atop 0}\right)$,
and
$|\{\mu\}_\beta\rangle$ indicates a state of the form
(\ref{bethe1}), with rapidities that satisfy the Bethe equations
(\ref{bethe2}). We call such a state a {\it Bethe eigenstate}.

\subsection{Scalar products}

It is possible to define a space dual to that of Equation
(\ref{bethe1})

\begin{equation}
\langle\{\lambda\}| = \langle 0| \, C(\lambda_1) \ldots C(\lambda_N)
\label{dual}
\end{equation}

\noindent where $\langle 0| = \otimes^{M} (1\ 0)$, and take
scalar products of the two

\begin{equation}
\langle \{ \lambda \} | \{ \mu \} \rangle
=
\langle 0|
\, C(\lambda_1) \ldots C(\lambda_N)\, B(\mu_1)\ldots B(\mu_N)|0
\rangle
\label{sp1}
\end{equation}

The scalar product (\ref{sp1}) can be written as a complicated
sum over a product of two determinants, \cite{korepin-book}.
Simpler expressions exist when the variables $\{\lambda\}$ 
and/or $\{\mu\}$ satisfy the Bethe equations. 

\subsection{Bethe scalar products} Let us consider the case 
when the set $\{\lambda\}$ are free variables, and the set 
$\{\mu\}$ satisfy the Bethe equations. We will denote the
latter set by $\{\mu\}_{\beta}$. Scalar products of the form
$\langle \{ \lambda \} | \{ \mu \}_\beta \rangle$ play 
a central role in the calculation of correlation functions
\cite{kitanine}.

\subsection{Determinant expressions for the Bethe scalar products}

Following \cite{slavnov}, the Bethe scalar product (\ref{sp1}) 
can be written as 

\begin{equation}
\langle \{ \lambda \} | \{ \mu \}_\beta \rangle
=
\frac{[\gamma]^N \prod_{i,j=1}^{N}[\lambda_i-\mu_j+\gamma]}
{\prod_{1\leq i<j \leq N}[\lambda_i-\lambda_j] [\mu_j-\mu_i]}
\prod_{k=1}^{N}
d(\lambda_k) \, d(\mu_k)
\, \det
\Omega
\label{slavnov}
\end{equation}

\noindent where the components of the $N\times N$ matrix
$\Omega$ are

\begin{multline}
\Omega_{ij}
=
\frac{1}{[\lambda_i-\mu_j] [\lambda_i-\mu_j+\gamma]}
-
\frac{(-)^{N}}{[\mu_j-\lambda_i] [\mu_j-\lambda_i+\gamma]}
\frac{a(\lambda_i)}{d(\lambda_i)}
\prod_{k=1}^{N}
\frac{[\mu_k-\lambda_i+\gamma]}{[\lambda_i-\mu_k+\gamma]}
\label{components}
\end{multline}

Using the explicit form of the functions $a(\lambda)$ and
$d(\lambda)$ for the case of the XXZ model \cite{korepin-book},
we obtain

\begin{equation}
\langle \{ \lambda \} | \{ \mu \}_\beta \rangle
=
\frac{[\gamma]^N \prod_{i,j=1}^{N}[\lambda_i-\mu_j+\gamma]}
{\prod_{1\leq i<j \leq N}[\lambda_i-\lambda_j] [\mu_j-\mu_i]}
\prod_{k=1}^{N}\prod_{l=1}^{M}
\, [\lambda_k-\nu_l] \, [\mu_k-\nu_l]
\, \det
\Omega
\label{slavnov1}
\end{equation}

\noindent where the components of $\Omega$ now become

\begin{multline}
\Omega_{ij} = \frac{1}{[\lambda_i-\mu_j] [\lambda_i-\mu_j+\gamma]} -
\\
\frac{(-)^{N}}{[\mu_j-\lambda_i] [\mu_j-\lambda_i+\gamma]}
\prod_{k=1}^{M}
\frac{[\lambda_i-\nu_k+\gamma]}{[\lambda_i-\nu_k]}
\prod_{l=1}^{N}
\frac{[\mu_l-\lambda_i+\gamma]}{[\lambda_i-\mu_l+\gamma]}
%\label{components}
\end{multline}

We refer to \cite{slavnov} for details of the proof.

\section{The scalar product is a restricted KP $\tau$ function}
\label{scalar-product}

We now bring the determinant in Equation (\ref{slavnov1}) to the 
form of a restricted KP $\tau$ function. The first step is to 
rewrite it so that it is more clearly a trigonometric polynomial
in $\{\lambda\}$. This is done by absorbing the products in the
numerator of (\ref{slavnov1}) into the determinant, to obtain

\begin{equation}
\langle \{ \lambda \} | \{ \mu \}_\beta \rangle
=
\frac{[\gamma]^N \det \Omega'}
{\prod_{1 \leq i<j \leq N}[\lambda_i-\lambda_j] [\mu_j-\mu_i]}
\label{slavnov2}
\end{equation}

\noindent where the components of $\Omega'$ are

\begin{multline}
\label{omega-prime}
\Omega'_{ij}
=
\frac{\prod_{k=1}^{M} [\mu_j-\nu_k]}{[\lambda_i-\mu_j]}\times
\\
\ll
\prod_{k=1}^{M}
[\lambda_i-\nu_k]
\prod_{l\not=j}^{N}[\lambda_i-\mu_l+\gamma]
+(-)^N
\prod_{k=1}^{M}
[\lambda_i-\nu_k+\gamma]
\prod_{l\not=j}^{N}[\mu_l-\lambda_i+\gamma]
\rr
\end{multline}

Using the Bethe equations (\ref{bethe2}) to rewrite the factor 
$\prod_{k=1}^{M} [\mu_j-\nu_k]$ in the numerator of the right 
hand side of Equation (\ref{omega-prime}), and extracting an 
overall factor from the resulting determinant, we obtain

\begin{equation}
\langle \{ \lambda \} | \{ \mu \}_\beta \rangle
=
\frac{
[\gamma]^N \det \Omega''
}{
\ll
\prod_{1\leq i<j \leq N}[\lambda_i-\lambda_j] [\mu_j-\mu_i]
\rr
\prod_{i \not= j} [\mu_i-\mu_j+\gamma] }
\label{slavnov3}
\end{equation}

\noindent where the components of the $N\times N$ matrix
$\Omega''$ are

\begin{multline}
\Omega''_{ij}
=
\frac{(-)^N}{[\lambda_i-\mu_j]}
\ll
\prod_{k=1}^{M}
[\lambda_i-\nu_k+\gamma]
[\mu_j-\nu_k]
\prod_{l\not=j}^{N}
[\mu_l-\lambda_i+\gamma]
[\mu_j-\mu_l+\gamma]
\right.
\\
-
\left.
\prod_{k=1}^{M}
[\lambda_i-\nu_k]
[\mu_j-\nu_k+\gamma]
\prod_{l\not=j}^{N}
[\lambda_i-\mu_l+\gamma]
[\mu_l-\mu_j+\gamma]
\rr
\end{multline}

Equation (\ref{slavnov3}) is a trigonometric polynomial in
the variables $\{\lambda\}$, since all poles in the expression
are removable. This can be seen as follows. The poles at
$\lambda_i=\lambda_j$,
$i\not= j$ are canceled by the zeros in the determinant when
two rows become equal. Also, the poles at $\lambda_i=\mu_j$
within $\Omega''_{ij}$ are canceled by corresponding zeros
in the numerator. We now come to the statement of our result.

Consider the normalized scalar product

\begin{equation}
\ll
\prod_{i=1}^{N} e^{(M-1)(\lambda_i+\mu_i)}
\prod_{j=1}^{M}e^{2N\nu_j}
\rr
\langle \{ \lambda \} | \{ \mu \}_\beta
\rangle
\end{equation}

\noindent and make the change of variables 

\begin{equation}
\label{set}
\{
e^{2\lambda_i}, 
e^{2    \mu_i},
e^{2    \nu_i},
e^{    \gamma}
\}
\rightarrow
\{
x_i, y_i, z_i, q
\}
\end{equation}

\noindent to obtain

\begin{equation}
\ll
\prod_{i=1}^{N} e^{(M-1)(\lambda_i+\mu_i)}
\prod_{j=1}^{M}e^{2N\nu_j}
\rr
\langle \{ \lambda \} | \{ \mu \}_\beta \rangle
\rightarrow
\\
\langle
\{x\}|\{y\}_\beta
\rangle
\end{equation}

\noindent where 

\begin{equation}
\langle
\{x\}|\{y\}_\beta
\rangle
=
\frac{
(q-q^{-1})^N
\det\mathbf{\Omega}
}{
\ll
\prod_{1\leq i < j \leq N}\left(x_i-x_j\right)\left(y_j-y_i\right)
\rr
\prod_{i\not= j}\left(y_i q-y_j q^{-1}\right)
}
\label{theorem}
\end{equation}

\noindent and the components of $\mathbf{\Omega}$ are

\begin{multline}
\mathbf{\Omega}_{ij} = \frac{1}{x_i-y_j}
\ll
\right.
\prod_{k=1}^{M}
\left(x_i -z_k \right)
\left(y_j q-z_k q^{-1}\right)
\prod_{l\not=j}^{N}
\left(x_i q-y_l q^{-1}\right)
\left(y_j q^{-1}-y_l q\right)
\\
-
\prod_{k=1}^{M}
\left(x_i q-z_k q^{-1}\right)
\left(y_j -z_k \right)
\prod_{l\not=j}^{N}
\left(x_i q^{-1}-y_l q\right)
\left(y_j q -y_l q^{-1}\right)
\left.
\rr
\label{omega}
\end{multline}

\bigskip
\noindent
{\bf Lemma 3.}
\ \ \
Let $\{y\}_{\beta}$ be a set of $N$ variables that correspond
to the set $\{\mu\}_{\beta}$ that satisfy the Bethe equations 
(\ref{bethe2}), then
$\langle \{x\}|\{y\}_\beta\rangle$
as defined in Equation (\ref{theorem}),
is a restricted KP $\tau$ function in the variables $\{x\}$,
where $x_i = e^{2 \lambda_i}$.

\medskip
\noindent
{\bf Proof.}
\ \ \ The proof is a corollary of Lemma {\bf 1} and Lemma {\bf 2}.
As we are considering the expression in Equation (\ref{theorem})
to be a restricted KP $\tau$ function in the $\{x\}$ variables,
we treat all pre-factors {\it not} involving $\{x\}$ as constants,
and we need only show that
$
\frac{
\det{ \mathbf{\Omega}}
}{
\prod_{1\leq i < j \leq N} \left( x_i - x_j \right)}$ is in the
form of a restricted KP $\tau$ function. We start by writing

\begin{equation}
\label{proof1}
\mathbf{\Omega}_{ij}
=
\frac{\sum_{k=1}^{N+M} x_i^{k-1} \rho_{kj}}{x_i-y_j}
\end{equation}

\noindent where the coefficients $\rho_{kj}$ are given by

\begin{multline}
\rho_{kj}
=
\ll
\prod_{m=1}^{M}
\left(y_j q -z_m q^{-1} \right)
\rr
\ll
\prod_{n\not=j}^{N}
\left(y_j - y_n q^2 \right)
\cdot
e_{(M+N-k)} \{-\widehat{y}_j q^{-2} \}\{-z \}
\rr
\\
\label{proof2}
-
\ll
\prod_{m=1}^{M}
\left(y_j q -z_m q \right)
\rr
\ll
\prod_{n\not=j}^{N}
\left(y_j  - y_n q^{-2}\right)
\cdot
e_{(M+N-k)} \{ -\widehat{y}_j q^2 \}\{-z q^{-2} \}
\rr
\end{multline}

\noindent and $e_k \{\widehat{y}_j\} \{z\}$ is the $k$-th
elementary symmetric polynomial (\ref{e-s}) in the set of
variables $\{y\}\cup \{z\}$ with the omission of $y_j$. From 
Equation (\ref{omega}), the numerator of $\mathbf{\Omega}_{ij}$ 
vanishes for $x_i=y_j$, hence the right hand side of Equation 
(\ref{proof1}) is a polynomial in $\{x\}$

\begin{equation}
\label{proof3}
\mathbf{\Omega}_{ij}
=
\sum_{k=1}^{N+M-1}  
x_i^{k-1}
\ll 
- 
\sum_{l=1}^{k} 
y_j^{l-k-1}
\,
\rho_{lj} 
\rr
\end{equation}

Combining the results of Equation (\ref{proof1}--\ref{proof3}), 
we obtain

\begin{equation}
\mathbf{\Omega}_{ij} = 
\sum_{k=1}^{N+M-1} 
x_i^{k-1} \, \kappa_{kj},
\quad
\textrm{where}
\quad
\kappa_{kj} = -\sum_{l=1}^{k} y_j^{l-k-1} \, \rho_{lj} 
\label{needed}\end{equation}

The coefficients $\kappa_{kj}$ do not depend on $\{x\}$ or the
row-index $i$, and the required result follows from Equation
(\ref{lemma-matrix}).

\subsection{General scalar products are not $\tau$ functions}

We have verified by explicit checks of nontrivial cases that 
$\langle \{\lambda\}|\{\mu\}\rangle$, where neither sets of
rapidities satisfy the Bethe equations, is {\it not\/}
a restricted KP $\tau$ function.

\subsection{The domain wall partition function as a special case}

The Bethe scalar product $\langle \{\lambda\}|\{\mu\}_\beta\rangle$ 
contains two sets of rapidities $\{\lambda\}$ and $\{\mu\}_\beta$, 
each of cardinality $N$, and a set of inhomogeneities $\{\nu\}$
of cardinality $M$. In the particular case when $M=N$, the scalar
product reduces to a product of two domain wall partition
functions \cite{korepin-book}

\begin{multline}
\label{PF}
\langle \{\lambda\}|\{\mu\}_\beta\rangle
=
\langle 0|C(\lambda_1)\ldots C(\lambda_N)|1\rangle
\langle 1|B(\mu_1) \ldots B(\mu_N)|0\rangle
\\
=
Z_N
\ll
\{\lambda\},\{\nu\}
\rr
Z_N
\ll
\{\mu\}_\beta,\{\nu\}
\rr
\end{multline}

\noindent where 
$|1\rangle  = \otimes^N \ll 0 \atop 1 \rr$ and 
$\langle 1| = \otimes^N (   0 \     1)$. 
Multiplying both sides of Equation (\ref{PF}) by
$\prod_{i=1}^{N} e^{(N-1)(\lambda_i+\mu_i)}e^{2N \nu_i}$ 
and using the change of variables in Equation (\ref{set}), 
we obtain

\begin{equation}
\langle \{x\}|\{y\}_\beta \rangle
=
Z_N
\ll
\{x\},\{z\}
\rr
Z_N
\ll
\{y\}_\beta,\{z\}
\rr
\label{PF2}
\end{equation}

{}From Lemma {\bf 3}, $\langle \{x \}|\{y\}_\beta\rangle$
on the right hand side of Equation (\ref{PF2}) is a restricted 
KP $\tau$ function in the variables $\{x\}$. Considering
$Z_N \ll \{y\}_\beta, \{z\}\rr$
to be a multiplicative constant, we conclude that
$Z_N \ll \{x\},       \{z\}\rr$
is a restricted KP $\tau$ function in the variables $\{x\}$. This is
the result obtained in \cite{fwz}. 

\section{Fermionic expectation values}
\label{fermionic}

In this section we reconsider the identity given in Equation
(\ref{lemma-matrix}) and express it as a vacuum expectation 
value of charged free fermions, with
restricted time variables. The derivation of the result
proceeds {\it mutatis mutandis} as that of an analogous
result for the domain wall partition function discussed
in detail in \cite{fwz}, hence we only give the final results 
and refer to \cite{fwz} for the details.

\subsection{Charged free fermions and vacuum states}

The free fermion operators $\{\psi_{n}, \psis_{n}\}$,
$n \in \Z$, with charges $\{+1, -1\}$
and energies $n$, satisfy the anti-commutation relations
\begin{equation}
\left.
\begin{array}{l}
\left[  \psi_m,  \psi_n \right]_{+} = 0 \\ \\
\left[ \psis_m, \psis_n \right]_{+} = 0 \\ \\
\left[  \psi_m, \psis_n \right]_{+}= \delta_{m,n}
\end{array}
\right\}\ \forall \ m, n \in \Z
\label{aa}
\end{equation}

The vacuum states $\langle 0|$ and $|0\rangle$ are defined 
by the actions

\begin{equation}
\left.
\begin{array}{l}
\langle 0| \psi_{n} = \psi_m |0\rangle = 0,
\\ \\
\langle 0|\psis_{m} = \psis_n|0\rangle = 0,
\end{array}
\right\} \quad \forall\ m < 0,\ n \geq 0
\label{vacuum-states}
\end{equation}

\noindent and the inner product normalization
\begin{equation}
\langle 0 | 0 \rangle = 1
\label{norm}
\end{equation}

\subsection{Creation/annihilation operators and normal 
ordering} The annihilation operators are those which 
annihilate the vacuum state $|0\rangle$, that is 
$\psi_m |0\rangle = 0$, $m <   0$, and
$\psis_n|0\rangle = 0$, $n \ge 0$, while all other operators,
$\psi_m |0\rangle \neq 0$, $m \ge 0$, and
$\psis_n|0\rangle \neq 0$, $n  <  0$
are creation operators. The normal-ordered product is 
defined, as usual, by placing annihilation operators 
to the right of creation operators
\begin{equation}
:\psi_{i}\psis_{j}:\ = \psi_{i}\psis_{j}-\langle 0|
\psi_{i}\psis_{j}|0\rangle
\end{equation}

\subsection{The Heisenberg operators and the KP Hamiltonian}
The neutral bilinear operators

\begin{equation}
H_m = {\sum_{j \in \Z}:\psi_{j}\psis_{j+m}:},
\quad m \in \Z
\label{ak}
\end{equation}

\noindent together with the central element $1$ form
a Heisenberg algebra

\begin{equation}
\left[ H_m, H_n \right] = m \delta_{m + n, 0},
\ \forall \ m, n \in \Z
\label{al}
\end{equation}

\noindent and define the KP Hamiltonian

\begin{equation}
H\{t\} = \sum_{m = 1}^{\infty} t_m H_m
\end{equation}

\subsection{Boson-fermion correspondence}
The character polynomial $\chi_{\lambda} \{t\}$ can 
be generated as follows

\begin{equation}
\langle 0|
e^{H\{t\}}
\psis_{-b_1}  \cdots  \psis_{-b_d}
 \psi_{ a_d}  \cdots   \psi_{ a_1}
|0\rangle
=
(-1)^{b_1 + \cdots + b_d} \chi_{\lambda}\{t\}
\label{boson-fermion}
\end{equation}

\noindent where
$a_i$ is the (length $+1$) of the $i$-th horizontal part, and
$b_i$ is the length        of the $i$-th vertical   part in 
the Frobenius decomposition of $\lambda$ \cite{macdonald},
and we assume that 
$a_d < \cdots < a_1$ and
$b_d < \cdots < b_1$, 
where $d$ is the number of cells on the main diagonal of
$\lambda$.

\subsection{The Bethe scalar product as fermion expectation value}
Using the Cauchy-Binet formula to expand the right hand side 
of Equation (\ref{lemma-matrix}) in terms of Schur functions, 
we obtain

\begin{equation}
\det
\ll
\sum_{k=1}^{N+M-1}
h_{k-i}\{x\}
\, 
\kappa_{k,N-j+1}
\rr_{1 \leq i,j \leq N}
=
\sum_{\lambda \subseteq [(M-1)^{N}]}
 c_{\lambda} \, s_{\lambda}\{x\}
\label{lemma-matrix2A}
\end{equation}

\noindent where

\begin{equation}
c_{\lambda} =
\det
\ll
\kappa_{(\lambda_{(N-i+1)} +i), (N-j+1)}
\rr_{1 \leq i, j \leq N}
=
\det
\ll
\kappa_{(\lambda_{i} +N-i+1), j}
\rr_{1 \leq i, j \leq N}
\label{lemma-matrix3A}
\end{equation}

\bigskip
\noindent
{\bf Lemma 4.} The expansion in terms of Schur functions on the 
right hand side of Equation (\ref{lemma-matrix2A}) can be written 
as a fermion vacuum expectation value with restricted time variables

\begin{equation}
\sum_{\lambda \subseteq [(M-1)^{N}]}
c_{\lambda} s_{\lambda} \{x\}
=
c_{\phi}
\langle 0|
e^{H \{x\}}
e^{X_0 \{y\}_{\beta}} \dots e^{X_{M-2} \{y\}_{\beta}}
|0 \rangle
\label{lemma-matrix4}
\end{equation}

\noindent where 

\begin{equation}
H \{x\} = 
\sum_{n=1}^{\infty} \frac{1}{n} \sum_{i=1}^{N} x_i^n H_n
\end{equation}

\noindent and $X_j$ denotes the following sum of fermion bilinears

\begin{equation}
X_j \{y\}_{\beta} =
\sum_{k=1}^{N}
(-)^k d_{[j+1, 1^{(k-1)}]}
\psi^{*}_{-k} \psi_j,
\quad
j = \{0, \dots, M-2\}
\label{lemma-matrix5}
\end{equation}

\noindent with the coefficients $d_{\lambda} = c_{\lambda} / c_{\phi}$,
where $c_{\lambda}$ is defined in Equation (\ref{lemma-matrix3A}).

\noindent \textbf{Proof.} 
\ \ \ The proof of (\ref{lemma-matrix4}) is
identical to that of an analogous result for the domain wall
partition function in \cite{fwz}. The differences are that
{\bf 1.} The coefficients of the expansion of the initial
bosonic expression in terms of Schur functions are different,
but that does not change the proof, and 
{\bf 2.} The expression that turns out to be a restricted 
KP $\tau$ function is a polynomial of degree $N(M-1)$ 
rather than $N(N-1)$ in $N$ independent variables 
$\{x\}$, hence the expansion in terms of Schur functions 
is indexed by $\lambda \subseteq$ $[(M-1)^{N}]$ rather than 
$\lambda \subseteq [(N-1)^{N}]$.

Using Equations (\ref{theorem}) and (\ref{lemma-matrix4}), 
we obtain the following vacuum expectation expression for 
the scalar product

\begin{equation}
\langle \{x \} | \{y \}_{\beta} \rangle
=
\mathcal{N}
\, 
\langle 0| 
e^{H      \{x\}}
e^{X_0    \{y\}_{\beta}}
\dots 
e^{X_{M-2}\{y\}_{\beta}}
|0\rangle
\label{slav.15}
\end{equation}

\noindent where the multiplicative factor $\mathcal{N}$ is

\begin{equation}
\mathcal{N} =
\frac{
c_{\phi}
(q-q^{-1})^N
}{
\ll
\prod_{1 \le i< j \le N} (y_j-y_i)
\rr
\ll
\prod_{i \ne j} (y_i q-y_j q^{-1})
\rr
}
\end{equation}

\section{A correspondence}
\label{section-correspondence}

Consider the left state $\langle \{x\}|$ on the left hand side 
of Equation (\ref{slav.15}).
Given that $\{x\}$ are free variables,
this state is completely specified by 
{\bf 1.} The length $M$ of the spin chain,
{\bf 2.} The number $N$ of $C$ operators,
{\bf 3.} The $M$ inhomogeneity parameters $\{z\}$ 
and 
{\bf 4.} The crossing parameter $\gamma$ of the spin chain.
But all these data are available in the eigenstate
$|\{y\}_{\beta}\rangle$, hence the 
scalar product is encoded in and can be recovered from 
the right state $|\{y\}_{\beta}\rangle$.

Next, consider the left state 
$\langle 0| e^{H \{x\}}$
on the right hand side of Equation (\ref{slav.15}).
Once again, since $\{x\}$ are $N$ free variables, this inner
product is completely specified by data that are available in 
the right state
$e^{X_0    \{y\}_{\beta}} \ldots 
 e^{X_{M-2}\{y\}_{\beta}} |0\rangle$. 
{\it \lq Peeling off\rq\,} the left states which can be
reconstructed from the corresponding right states, we obtain 
the correspondence

\begin{equation}
\label{correspondence}
|\{\mu\}_{\beta}\rangle \longrightarrow
e^{X_0\{y\}_{\beta}} \ldots e^{X_{M-2}\{y\}_{\beta}} |0\rangle
\end{equation}

The left hand side of Equation (\ref{correspondence}) is
an XXZ Bethe eigenstate.
The right hand side is a point on Sato's Grassmannian
\cite{blue-book}. The correspondence in Equation
(\ref{correspondence}) is an injective map that assigns 
to each XXZ eigenstate a point on Sato's Grassmannian.

\section{Remarks}
\label{remarks}

The fact that quantum integrable models are related to classical 
integrable differential equations can be traced to the 
pioneering work \cite{wu}. Since then, quite a few results 
in along these lines have been obtained, particularly on the 
connection of the quantum Bose gas and the classical nonlinear 
Schr\"odinger equation, as reviewed in \cite{korepin-book}. 
Closer to the spirit of the present work is the result of 
\cite{iiks} that the quantum XXZ correlation 
functions at the free fermion point are $\tau$ functions of 
the Ablowitz-Ladik equation\footnote{We thank N~Kitanine for 
pointing this out to us.}. What is new in the present work 
is the result that the XXZ Bethe scalar product is a KP 
$\tau$ function for all values of the crossing parameter. 
The Bethe scalar 
products are basic building blocks of the XXZ correlation 
functions \cite{kitanine}, and we hope that our result will 
help in the current efforts to compute the latter and their 
asymptotics.

The correspondence between the Bethe eigenstates and points
on Sato's Grassmannian is reminiscent of Sklyanin's separation
of variables approach to quantum integrable models \cite{sklyanin},
where every solution to the Bethe equations, $\{y\}_{\beta}$, 
labels two objects,
{\bf 1.} A Bethe eigenstate, and
{\bf 2.} A function on the projective line with specific monodromy
properties. In our correspondence, we obtain a Bethe eigenstate
and a point on the Grassmannian. It is also reminiscent of the 
results of Mukhin {\it et al.} in the context of the Gaudin limit 
of the XXX model \cite{mukhin}. 

Recently, Nekrasov and Shatashvili obtained a correspondence between 
XXZ Bethe eigenstates and vacuum states of a 3-dimensional super 
Yang-Mills theory compactified on a circle 
\cite{nekrasov}. Combining 
the latter correspondence and ours points to a correspondence between 
a super Yang-Mills theory and points on Sato's Grassmannian. 
We hope to explore these issues in future publications.

\section*{Acknowledgments}
We wish to thank Professors M~Jimbo, N~Kitanine, V~E~Korepin, E~Mukhin, 
T~Shiota 
and K~Takasaki for useful remarks on this work and on related topics.
MW and MZ are supported by Australian Postgraduate Awards and by the
Department of Mathematics and Statistics, The University of Melbourne.

\end{document}